# THE CLIC MAIN LINAC ACCELERATING STRUCTURE

I. Wilson, W. Wuensch, CERN, Geneva, Switzerland

*Abstract*

This paper outlines the RF design of the CLIC (Compact Linear Collider) 30 GHz main linac accelerating structure and gives the resulting longitudinal and transverse mode properties. The critical requirement for multibunch operation, that transverse wakefields be suppressed by two orders of magnitude within 0.7 ns (twenty fundamental mode cycles), has been demonstrated in a recent ASSET experiment. The feasibility of operating the structure at an accelerating gradient of 150 MV/m for 130 ns has yet to be demonstrated. Damage of the internal copper surfaces due to high electric fields or resulting from metal fatigue induced by cyclic surface heating effects are a major concern requiring further study.

## 1 INTRODUCTION

A major effort has gone into developing an accelerating structure for the CLIC main linac that can operate with an average gradient of 150 MV/m, that maintains an acceptable short-range wakefield, and that has highly suppressed long-range transverse wake fields. The latter design criterion is a result of multibunching - a common feature among linear collider designs in which multiple-bunch trains are accelerated during each RF pulse in order to reach design luminosities ($10^{35}$ cm$^{-2}$ sec$^{-1}$ at 3 TeV for CLIC [1]) in a power-efficient way.

For CLIC, beam dynamics simulations indicate that the amplitude of the single-bunch transverse wakefield must decrease from its short-range value by a factor of 100 during the first 0.67 ns (which is the time between bunches in the train and which corresponds to twenty 30 GHz accelerating mode cycles) and that it must continue to decrease at least linearly for longer times [2].

A structure, called the TDS (Tapered Damped Structure), capable of such transverse wakefield suppression, has been developed [3]. The suppression is achieved primarily through damping. Each cell of the 150-cell TDS is damped by its own set of four individually terminated waveguides. This produces a Q of below 20 for the lowest dipole band. The waveguides have a cutoff frequency of 33 GHz which is above the 30 GHz fundamental but below all other higher-order modes. Cell and iris diameters are tapered along the length of the structure in order to induce a frequency spread in the dipole bands which is called detuning. Iris dimensions range from 4.5 to 3.5 mm. Detuning de-coheres the wakefield kicks further suppressing the transverse wakefield.

A first pass through the RF design of the TDS, for both fundamental mode performance and transverse wakefield suppression, has now been made. Certain weaknesses in the design, with regard to operation at very high gradients, have become apparent and are being addressed in a complete re-optimisation of the structure.

## 2. GEOMETRY AND FUNDAMENTAL MODE CHARACTERISTICS

The geometry of a TDS cell is shown in Figure 1. A full structure consists of 150 cells. Iris diameters (2a) vary linearly from 4.5 mm at the head of the structure to 3.5 mm at the tail. The relationship between iris diameter (2a) and cell diameter (2b) necessary to maintain a fundamental mode frequency of 29.985 GHz was obtained using HFSS. $Q$, $R'/Q$, and group velocity ($v_g$) as a function of $b$ were then calculated. The relationship between a and b is plotted in Figure 2 and $Q$, $R'/Q$, and $v_g$ are plotted in figure 3.

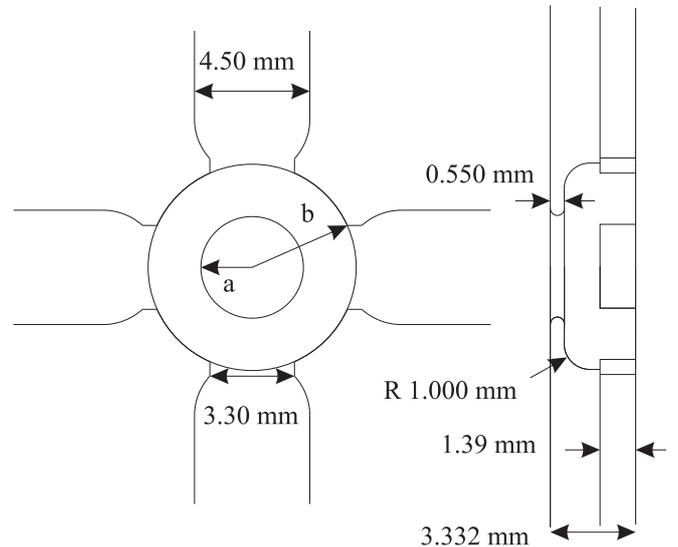

Figure1: Geometry of TDS cell and damping waveguides.

The nominal CLIC beam consists of a train of 0.6 nC bunches spaced by 20 cm. Operating the structure at an average accelerating gradient of 150 MV/m results in the loaded and unloaded accelerating gradient profiles plotted in Figure 4. The corresponding power flows are plotted in Figure 5.

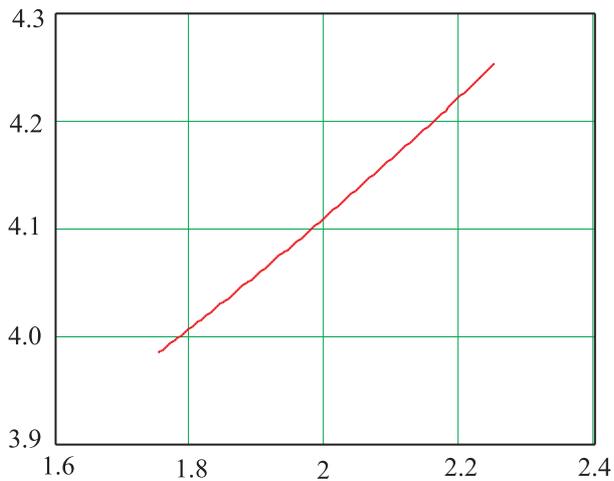

Figure 2: Relationship between 2b (y axis) and 2a (x axis) for correct fundamental mode frequency. Units are mm.

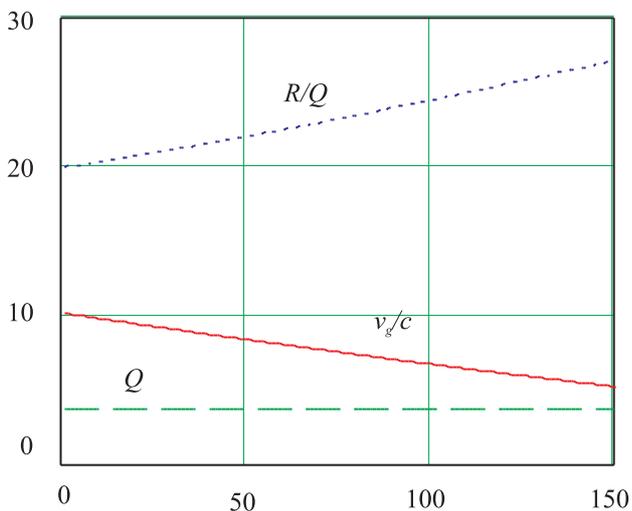

Figure 3: Fundamental mode characteristics as a function of cell number. $R'/Q$ is in kΩ/m, $v_g/c$ is a percentage. Q is nearly constant and has a value of 3600.

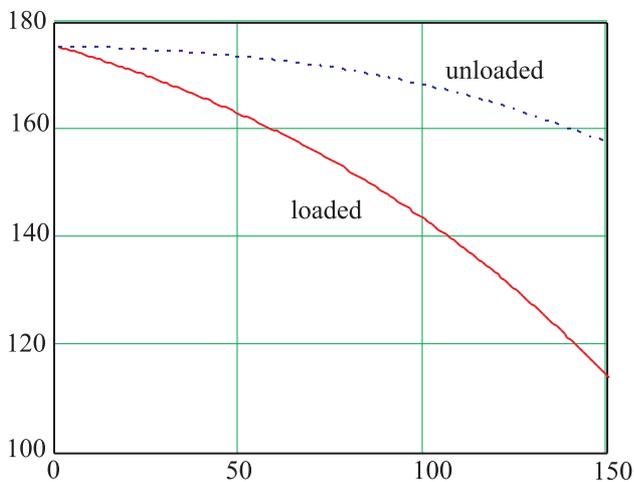

Figure 4: Loaded and unloaded accelerating gradient [MV/m] as function of cell number.

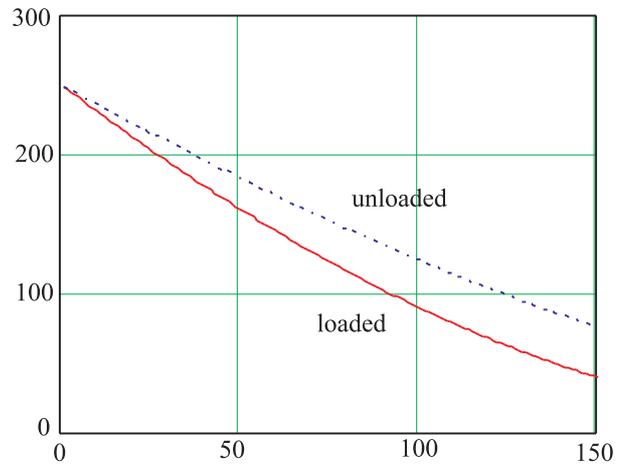

Figure 5: Loaded and unloaded power flows in units of MW as a function of cell number.

The highest surface field occurs in the first cell of the structure and has a value of 420 MV/m. For a 150 ns long pulse, a pulsed temperature rise of about 250 °C occurs in the cell wall between the waveguides.

## 3. DIPOLE MODE CHARACTERISTICS.

The damping waveguide terminating load is made from silicon carbide. The geometry of the load is shown in Figure 6. The principle behind this compact load, and the method by which the complex permittivity was obtained is described in reference [5]. The amplitude of the reflection coefficient, $|S_{11}|$, as a function of frequency was calculated using HFSS - see Figure 7. The transverse wakefield was then calculated using the double-band circuit model described in reference [4]. The result is plotted in Figure 8. The transverse wakefield suppression of the TDS and the validity of the tools used to model it was verified in an experiment in the ASSET facility [6] at SLAC. The calculated and measured wakefields are plotted in Figure 9.

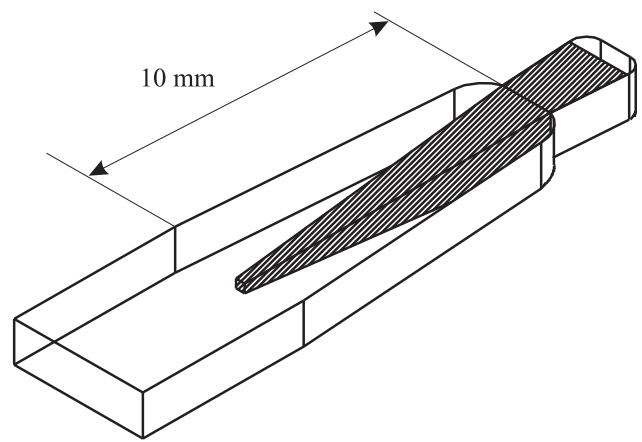

Figure 6: Geometry of the 30 GHz load.

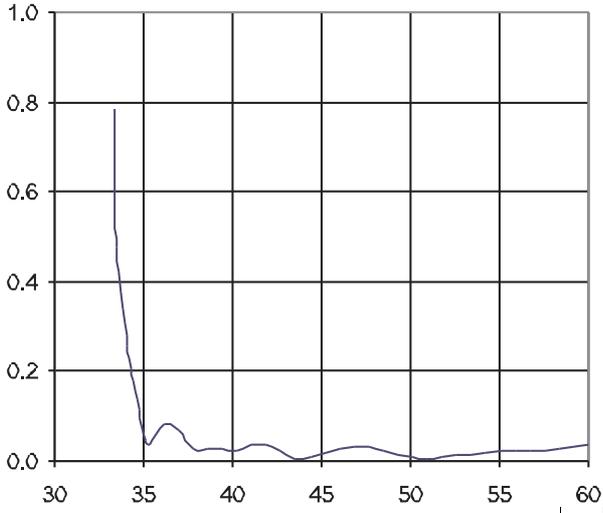

Figure 7: Amplitude of the reflection coefficient, $|S_{11}|$, as a function of frequency in units of GHz.

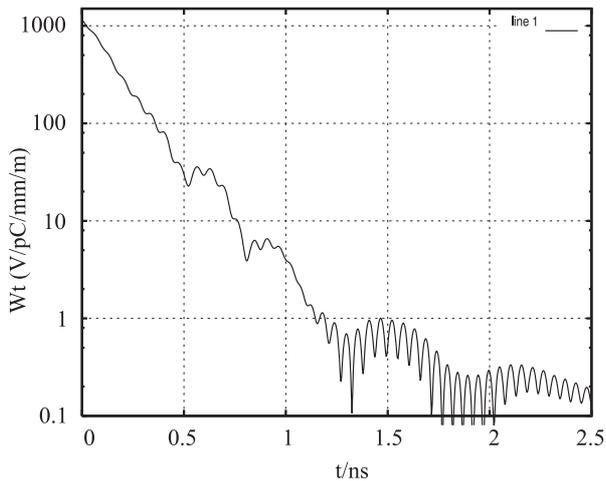

Figure 8: Computed transverse wakefield.

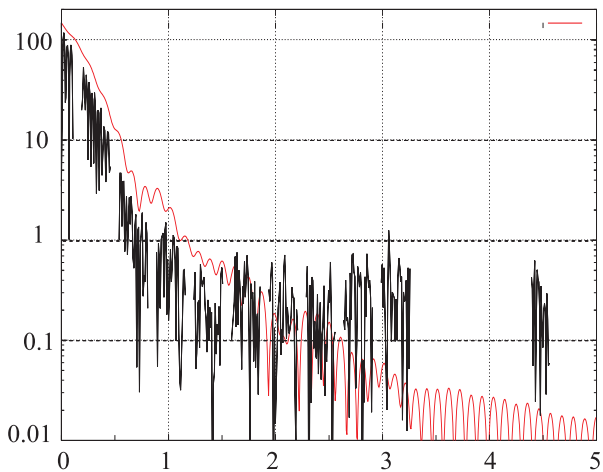

Figure 9: Results from the ASSET test. The amplitude of the transverse wake in units of V/(pC·mm·m) is plotted against time in ns. The test was made using a scaled 15 GHz structure.

## 4. CONCLUSIONS

A complete first pass has been made through the design of the TDS, for both the accelerating and the transverse modes. The transverse mode suppression is well within the specifications for CLIC and has been demonstrated experimentally.

The situation for the fundamental mode for an accelerating gradient of 150 MV/m is less favourable. The 420 MV/m peak surface electric field at the input of the structure is quite high, the feasibility of which must clearly be demonstrated.

The 250° pulsed surface heating of the cell walls is of even greater concern, since it is probably well above an acceptable value. The temperature rise may be reduced by optimising the cell geometry, in particular by decreasing both the thickness and width of the coupling iris. In this way the coupling of the damping waveguides to the dipole mode can be maintained while decreasing the perturbation of the fundamental mode currents.

An overall shift to smaller irises (dimension 2a) and fewer cells - thus maintaining the same RF to beam efficiency - would reduce both the peak surface electric field and the pulsed surface heating in the first cell. Beam dynamics simulations are needed to determine how much increase in the transverse wakefield due to the smaller irises can be tolerated.

## ACKNOWLEDGEMENTS

The authors would like to acknowledge their sincere and deep appreciation to Micha Dehler and Michel Luong, who have both made substantial contributions to the development of this structure. Special thanks are also extended to Erk Jensen for maintaining the circuit model.